\documentclass[12pt,a4paper]{article}
\usepackage[T1]{fontenc}
\usepackage[sc,osf]{mathpazo}
\usepackage{graphicx}
\usepackage{a4wide}  
\usepackage{latexsym,amsthm,amsfonts,amsmath,mathrsfs,amssymb}
\usepackage{booktabs} 
\usepackage{ifpdf}
\ifpdf
\usepackage[pdftex,unicode,implicit]{hyperref}
\hypersetup{%
  pdftitle    = {A gravitating Yang-Mills instanton}
  pdfkeywords = {Yang-Mills, instanton, supergravity, supersymmetry,
    event horizon, horizonless},
  pdfauthor   = {Pablo Cano Tomas Ortin, Pedro F. Ramirez},
  plainpages  = true,
  colorlinks  = true,
  citecolor   = blue,
  urlcolor    = red,
  linkcolor   = black
}
\newcommand{\hepth}[1]{{\tt
\href{http://www.arXiv.org/abs/hep-th/#1}{hep-th/#1}}}

\newcommand{\arxiv}[1]{{\tt arXiv:\href{http://www.arXiv.org/abs/#1}{#1}}}
\newcommand{\doi}[1]{{\tt DOI:\href{http://dx.doi.org/#1}{#1}}}

\else
  \usepackage[dvips]{graphicx}
  \usepackage[unicode,implicit]{hyperref}
  \newcommand{\hepth}[1]{{\tt hep-th/#1}}

  \newcommand{\arxiv}[1]{{\tt arXiv:#1}}
  \newcommand{\doi}[1]{{\tt DOI:#1}}

\fi
\makeatletter
\@addtoreset{equation}{section}
\makeatother

\begin{document}

\begin{flushright}
\small
IFT-UAM/CSIC-17-019\\
\texttt{arXiv:1704.00504 [hep-th]}\\
April 3\textsuperscript{rd}, 2017\\
\normalsize
\end{flushright}

\vspace{1.5cm}

\begin{center}

{\LARGE {\bf {A gravitating Yang-Mills instanton}}}
 
\vspace{1.5cm}

\renewcommand{\thefootnote}{\alph{footnote}}
{\sl\large Pablo A.~Cano}\footnote{E-mail: {\tt pablo.cano [at] uam.es}},
{\sl\large Tom\'{a}s Ort\'{\i}n}\footnote{E-mail: {\tt Tomas.Ortin [at] csic.es}}
{\sl\large and Pedro F.~Ram\'{\i}rez}\footnote{E-mail: {\tt p.f.ramirez [at]  csic.es}},

\setcounter{footnote}{0}
\renewcommand{\thefootnote}{\arabic{footnote}}

\vspace{1.5cm}

{\it Instituto de F\'{\i}sica Te\'orica UAM/CSIC\\
C/ Nicol\'as Cabrera, 13--15,  C.U.~Cantoblanco, E-28049 Madrid, Spain}\\

\vspace{2.5cm}


{\bf Abstract}

\end{center}

\begin{quotation}
  We present an asymptotically flat, spherically symmetric, static, globally
  regular and horizonless solution of SU$(2)$-gauged $\mathcal{N}=1,d=5$
  supergravity. The SU$(2)$ gauge field is that of the BPST instanton. We
  argue that this solution, analogous to the global monopoles found in $d=4$
  $\mathcal{N}=2$ and $\mathcal{N}=4$ gauged supergravities, describes the
  field of a single string-theory object which does not contribute to
  the entropy of black holes when we add it to them and show that it is,
  indeed, the dimensional reduction on T$^{5}$ of the gauge 5-brane. We
  investigate how the energy of the solution is concentrated as a function of
  the instanton's scale showing that it never violates the hoop conjecture
  although the curvature grows unboundedly in the zero scale limit.
\end{quotation}

\newpage
\pagestyle{plain}



\section*{Introduction}

Fueled by the research on theories of elementary particles and fundamental
fields (Yang-Mills, Kaluza-Klein, Supergravity, Superstrings...), over the
last 30 years, the search for and study of solutions of theories of gravity
coupled to fundamental matter fields (scalars and vectors in $d=4$ and
higher-rank differential forms in higher dimensions) has been enormously
successful and it has revolutionized our knowledge of gravity itself. Each new
classical solution to the Einstein equations (vacua, black holes, cosmic
strings, domain walls, black rings, black branes, multi-center solutions...)
sheds new light on different aspects of gravity and, often, on the underlying
fundamental theories. For instance, although the string effective field
theories (supergravities, typically) only describe the massless modes of
string theory, it is possible to learn much through them about the massive
non-perturbative states of the fundamental theory because they appear as
classical solutions of the effective theories.\footnote{See, \textit{e.g.},
  Refs.~\cite{Duff:1994an,Stelle:1998xg,Ortin:2015hya}.} Beyond this, there is
a definite program in the quest to construct horizonless \textit{microstate
  geometries} as classical solutions of Supergravity theories
\cite{Bena:2007kg, Bena:2013dka}. When interpreted within the context of the
fuzzball conjecture \cite{Mathur:2005zp}, these geometries have been proposed
to correspond to the classical description of black hole
microstates. Therefore, in the best case scenario, it might be possible to
find a large collection ($\sim e^S$) of microstate geometries with the same
asymptotic charges as a particular black hole, and, furthermore, to identify
explicitly their role in the ensemble of black-hole microstates. See
Refs.~\cite{Bena:2014qxa,Bena:2016ypk} for recent progress in that direction.
  
Apart from the fact that they describe gravity, one of the most interesting
features of string theories is that their spectra include non-Abelian
Yang-Mills (YM) gauge fields. This aspect is crucial for their use in BSM
phenomenology but has often been neglected in the search for classical
solutions of their effective field theories, specially in lower dimensions,
which have been mostly focused on theories with Abelian vector fields and
with, at most, an Abelian gauging. Thus, the space of extremal (supersymmetric
and non-supersymmetric, spherically-symmetric and multi-center) black-hole
solutions of 4- and 5-dimensional ungauged supergravities has been
exhaustively explored and progress has been made in the Abelian gauged case,
motivated by the AdS/CFT correspondence, but the non-Abelian case has drawn
much less attention in the string community and, correspondingly, there are
just a few solutions of the string effective action (and of supergravity
theories in general) with non-Abelian fields in the literature.

One of the main reasons for that is the intrinsic difficulty of solving the highly
non-linear equations of motion. This difficulty, however, has not prevented
the General Relativity community from attacking the problem in simpler
theories such as the Einstein-Yang-Mills (EYM) or Einstein-Yang-Mills-Higgs
(EYMH) theories, although it has prevented them from finding analytical
solutions: most of the genuinely non-Abelian solutions\footnote{That is,
  solutions whose non-Abelian fields cannot be rotated into Abelian ones using
  (singular or non-singular) gauge transformations. When they can be rotated
  into a purely Abelian one, it is often referred to as an ``Abelian
  embedding''.} are known only numerically.\footnote{The most complete review
  on non-Abelian solutions containing the most relevant developments until
  2001 is Ref.~\cite{Volkov:1998cc} complemented with the update
  Ref.~\cite{Galtsov:2001myk}. Ref.~\cite{Winstanley:2008ac} reviews the
  anti-De Sitter case. A more recent but less exhaustive review is
  Ref.~\cite{Volkov:2016ehx}, although it omits most of the non-Abelian
  solutions found recently in the supergravity/superstring context.} Another
reason is that non-Abelian YM solutions are much more difficult to understand
than the Abelian ones (specially when they are known only numerically): in the
Abelian case we can characterize the electromagnetic field of a black hole,
say, by its electric and magnetic charge, dipoles and higher multipoles. In
the non-Abelian case the fields are usually characterized by topological
invariants or constructions such as t'~Hooft's magnetic monopole charge.

In general, the systems studied by the GR community (the EYM or EYMH theories
in particular) are not part of any theory with extended local supersymmetry (a
$\mathcal{N}>1$ supergravity with more than 4 supercharges)\footnote{The supersymmetric solutions of
  $\mathcal{N}=1$ supergravity are massless (waves) or not asymptotically
  flat.} and, therefore, the use of supersymmetric solution-generating
techniques is not possible. One can, however, consider the minimal
$\mathcal{N}>1$ supergravity theories that include non-Abelian YM fields,
which are amenable to those methods. Some time ago we started the search for
supersymmetric solution-generating methods in $\mathcal{N}=2,d=4$
\cite{Hubscher:2008yz} and $\mathcal{N}=1,d=5$
\cite{Bellorin:2007yp,Bellorin:2008we,Bueno:2015wva,Meessen:2015enl}
Super-Einstein-Yang-Mills (SEYM) theories. The results obtained have allowed
to construct, for the first time (at least in fully analytical form), several
interesting supersymmetric solutions with genuine non-Abelian hair:
\textit{global monopoles} and extremal static black holes in 4
\cite{Meessen:2008kb,Bueno:2014mea,Meessen:2015nla} and 5 dimensions
\cite{Meessen:2015nla}, rotating black holes and black rings in 5 dimensions
\cite{Ortin:2016bnl}, non-Abelian 2-center solutions in 4 dimensions
\cite{Bueno:2014mea} and the first non-Abelian microstate geometries
\cite{Ramirez:2016tqc}.

Many of the black-hole solutions found by these methods can be embedded in
string theory and, in that framework, one can try to address the microscopic
interpretation of their entropy, which seems to have relevant contributions
from the non-Abelian fields, even though, typically, they decay so fast at
infinity that they do not seem to contribute to the mass. Following the
pioneer's route \cite{Strominger:1996sh,Maldacena:1996ky} requires an
understanding of the stringy objects (D-branes etc.) that contribute to the 4-
and 5-dimensional solutions' charges. Furthermore, the interpretation of the
non-Abelian microstate geometries would benefit from the knowledge of their
stringy origin. In this paper, as a previous step towards the microscopic
interpretation of the 5-dimensional non-Abelian black holes' entropy which we
will undertake in a forthcoming publication \cite{Cano:2017qrq}, we identify the
elementary component of the simplest, static, spherically symmetric,
non-Abelian 5-dimensional black hole that carries all the non-Abelian
hair. The solution that describes this component turns out to be
asymptotically flat, globally regular, and horizonless and the non-Abelian
field is that of a BPST instanton \cite{Belavin:1975fg} living in
constant-time hypersurfaces. Only a few solutions supported by elementary
fields with these characteristics are known analytically: the \textit{global
  monopoles} found in gauged $\mathcal{N}=4,d=4$ supergravity
\cite{Harvey:1991jr,Chamseddine:1997nm,Chamseddine:1997mc} and also in
$\mathcal{N}=2,d=4$ SEYM theories \cite{Hubscher:2008yz,Bueno:2014mea} whose
non-Abelian field is that of a BPS 't~Hooft-Polyakov monopole.

The simplest string embedding of this solution is in the Heterotic Superstring
and the 10-dimensional solution whose dimensional reduction over $T^{5}$ gives
this 5-dimensional \textit{global instanton} turns out to be the gauge 5-brane
found in Ref.~\cite{Strominger:1990et}. This is, therefore, the non-Abelian
ingredient present in the non-Abelian 5-dimensional black holes and rings
constructed in Refs.~\cite{Meessen:2015nla,Ortin:2016bnl}.

In what follows, we are going to derive the global instanton solution as a
component of the 5-dimensional non-Abelian black holes, we show that it is the
Heterotic String gauge 5-brane compactified on $T^{5}$ and we study the
dependence of the distribution of energy on the instanton's scale parameter,
showing that, no matter how small it is, there is never more energy
concentrated in a 3-sphere of radius $R$ than that of a
Schwarzschild-Tangerlini black hole of radius $R$. 

\section{The global instanton solution}
\label{sec-solution}

We are going to work in the context of the ST$[2,6]$ model of
$\mathcal{N}=1,d=5$ supergravity (which is a model with 5 vector
supermultiplets) with an SU$(2)$ gauging in the $I=3,4,5$ sector.  This theory
is briefly described in Appendix~\ref{sec-thetheory} and the solution-generating technique that allows us to construct
timelike supersymmetric solutions of this theory with one isometry is
explained in Appendix~\ref{sec-susysolutions}.

Our goal is to construct the minimal non-singular solution that includes in
the SU$(2)$ sector the following solution of the Bogomol'nyi equations

\begin{equation}
\Phi^{A}
=
\frac{1}{g_{4}r(1+\lambda^{2}r)} \frac{x^{A}}{r}\, ,
\hspace{1cm}
\breve{A}^{A}{}_{B}
=
\varepsilon^{A}{}_{BC}\frac{1}{g_{4}r(1+\lambda^{2}r)} \frac{x^{C}}{r}\, .  
\hspace{1cm}
r^{2} \equiv x^{s}x^{s}\, ,
\end{equation}

%

This solution describes a \textit{coloured monopole}
\cite{Meessen:2008kb,Meessen:2015nla}, one of the singular solutions found by
Protogenov \cite{Protogenov:1977tq}. Observe that this solution is written in
terms of the 4($=1+3$)-dimensional Yang-Mills coupling constant $g_{4}$. As shown in \cite{Bueno:2015wva}, the
4-dimensional Euclidean SU$(2)$ gauge field $\hat{A}^{A}$ that one obtains via
Eq.~(\ref{eq:Kromheimer}) for $H=1/r$ is the BPST instanton
\cite{Belavin:1975fg}, which justifies our choice. Using
the 4-dimensional radial coordinate $\rho^{2}=4r$, the 5-dimensional
Yang-Mills coupling constant $g_{4}=-2\sqrt{6}g$, and renaming
$4\lambda^{-2}=\kappa^{2}$ (the instanton scale parameter) it takes the
form\footnote{Our conventions for the SU$(2)$ gauge fields are slightly
  different from the ones used in Refs.~\cite{Meessen:2015enl,Ortin:2016bnl}:
  in this paper the generators satisfy the algebra $[T_{A},T_{B}]=
  +\epsilon_{ABC}T_{C}$ (which is equivalent to changing the sign of all the
  generators), and the gauge field strength is defined by $F= dA + gA\wedge
  A$. The left- and right-invariant Maurer-Cartan 1-forms $v_{L,R}$ have the
  same definitions, but the overall signs of the components are different, as
  a consequence of the change of sign in the generators $T_{A}$.}

\begin{equation}
\hat{A}^{A} = \frac{\kappa^{2}}{g(\rho^{2} +\kappa^{2})}v_{R}^{A}\, ,  
\end{equation}

\noindent
where the $v_{R}^{A}$ are the three SU$(2)$ left-invariant Maurer-Cartan
1-forms.

Let us now consider the ungauged sector. As it is well known, 5-dimensional
asymptotically-flat, static, regular black holes need to be sourced by at least three charges,
associated to three different kind of branes. A popular example is the D1D5W
black hole considered by Strominger and Vafa in
Ref.~\cite{Strominger:1996sh}. The corresponding solution of the
(supergravity) effective action is expressed in terms of three independent harmonic
functions. In the basis that we are using, these functions are $L_{0,1,2}$,
where the last two will be used in the the combinations $L_{\pm}= L_{1}\pm
L_{2}$ in order to make contact with the literature.

Thus, we take\footnote{The simplest 5-dimensional non-Abelian black hole
  constructed in Ref.~\cite{Meessen:2015enl} has $L_{2}=0$, or $L_{+}=L_{-}$
  and, therefore, it has three Abelian charges as well, but two of them are
  equal, which obscures the interpretation of the solution from the string
  theory point of view.}

\begin{equation}
L_{0,\pm} = B_{0,\pm} + q_{0,\pm}/\rho^{2}\, ,
\end{equation}

\noindent
and we will assume that all the constants are positive.  

This choice gives a static solution ($\hat{\omega}=0$, see the appendices for
more information) with the following active fields function

\begin{equation}
\label{eq:3chargebh1}
\begin{array}{rcl}
ds^{2} 
& = & 
\hat{f}^{2}dt^{2}-\hat{f}^{-1}(d\rho^{2}+\rho^{2}d\Omega_{(3)}^{2})\, ,
\\
& & \\
A^{0} 
& = &
-\tfrac{1}{\sqrt{3}}{\displaystyle\frac{1}{\tilde{L}_{0}}}dt\, ,
\hspace{1cm}
A^{1}\pm A^{2} 
= 
-\tfrac{2}{\sqrt{3}}{\displaystyle\frac{1}{L_{\pm}}} dt\, , 
\hspace{1cm}
A^{A} 
=
{\displaystyle
\frac{\kappa^{2}}{g(\rho^{2} +\kappa^{2})}v_{R}^{A}\, ,  
}
\\
& & \\
e^{2\phi}
& = &
2 \dfrac{\tilde{L}_{0}}{L_{-}}\, ,
\hspace{1cm}
k
=
(3\hat{f}L_{+})^{3/4}\, ,
\end{array}
\end{equation}

\noindent 
where the metric function $\hat{f}$ is given by

\begin{equation}
\label{eq:3chargebh2}
\hat{f}^{-1}
=
\left\{
\tfrac{27}{2}\tilde{L}_{0}L_{+}L_{-}
\right\}^{1/3}\, ,
\end{equation}

\noindent
and we have defined the combination

\begin{equation}
\tilde{L}_{0} \equiv  L_{0}- \tfrac{1}{3}\rho^{2}\Phi^{2}\, ,
\,\,\,\,\,\,\,
\mbox{and}
\,\,\,\,\,\,\,
\Phi^{2} 
\equiv 
\Phi^{A}\Phi^{A} 
= 
\frac{2 \kappa^{4}}{3g^{2}\rho^{4}(\rho^{2} +\kappa^{2})^{2}}\, .   
\end{equation}

The normalization of the metric at spatial infinity demands
$\frac{27}{2}B_{0}B_{+}B_{-}= 1$ and we can express the three integration
constants $B$ in terms of the values of the 2 scalars at infinity:

\begin{equation}
B_{0} = \tfrac{1}{3} e^{\phi_{\infty}}k_{\infty}^{-2/3}\, ,
\hspace{1cm}
B_{-} = \tfrac{2}{3} e^{-\phi_{\infty}}k_{\infty}^{-2/3}\, ,
\hspace{1cm}
B_{+} = \tfrac{1}{3} k_{\infty}^{4/3}\, ,
\end{equation}

\noindent
and the metric takes the form 

\begin{equation}
\hat{f}^{-1}
=
\left\{
(\tilde{L}_{0}/B_{0})\, (L_{+}/B_{+})\, (L_{-}/B_{-})
\right\}^{1/3}\, ,
\end{equation}

\noindent
where

\begin{equation}
  \begin{array}{rcl}
\tilde{L}_{0}/B_{0}
& = &
{\displaystyle
1
+
\frac{2 e^{-\phi_{\infty}}k_{\infty}^{2/3}}{3g^{2}}
\frac{\rho^{2} +2\kappa^{2}}{(\rho^{2} +\kappa^{2})^{2}}
+
3e^{-\phi_{\infty}}k_{\infty}^{2/3}\left(q_{0}- \frac{2}{9g^{2}}\right)\frac{1}{\rho^{2}}
\, ,
}
\\
& & \\
L_{-}/B_{-}
& = &
1+3e^{\phi_{\infty}}k_{\infty}^{2/3}q_{-}/(2\rho^{2})\, , 
\\
& & \\
L_{+}/B_{+}
& = &
1+3 k_{\infty}^{-4/3}q_{+}/\rho^{2}\, .  
\end{array}
\end{equation}

If $\tilde{q}_{0}\equiv q_{0}-\frac{2}{9g^{2}}>0$ and $q_{\pm}\neq 0$ there
is a regular event horizon with entropy

\begin{equation}
S =\frac{\pi^{2}}{2G_{N}^{(5)}}  \sqrt{ (3\tilde{q}_{0})\, (3q_{-}/2)\, (3q_{+})}\, .  
\end{equation}

The mass, however, depends on $q_{0}$, not on $\tilde{q}_{0}$

\begin{equation}
\label{eq:mass}
M
= 
\frac{\pi}{4G_{N}^{(5)}}  
\left[
e^{-\phi_{\infty}}k_{\infty}^{2/3}(3q_{0})
+e^{\phi_{\infty}}k_{\infty}^{2/3}(3q_{-}/2) 
+ k_{\infty}^{-4/3}(3q_{+}) \right]\, ,
\end{equation}

\noindent
so that the Yang-Mills fields only appear to be relevant in the near-horizon
region, a behavior also observed in 4-dimensional \textit{colored} black holes
Refs.~\cite{Meessen:2008kb,Meessen:2015nla}. Explaining this behavior and
finding a stringy microscopic interpretation for the entropy of these black
holes will be the subject of a forthcoming paper \cite{Cano:2017qrq}.  

One of the main ingredients needed to reach that goal is the list of
elementary components (branes, waves, KK monopoles...) of the black-hole
solution. In the Abelian case, these are typically associated to the harmonic
functions in which the brane charges occur as coefficients of the $1/\rho^{2}$
terms (in 5 dimensions) and these are the charges that appear in the entropy
formula. In the present case $\tilde{L}_{0}/B_{0}$ has a term which is finite in the $\rho\rightarrow 0$ limit and another term,
proportional to $\tilde{q}_{0}$, which goes like $1/\rho^{2}$ in that limit, as
an ordinary Abelian contribution would. The presence of the finite term
suggests the presence of a solitonic brane which does not contribute to the
entropy. 

In order to identify this brane we set $\tilde{q}_{0}=q_{\pm}=0$ in the above
solution (but $q_{0}=\frac{2}{9g^{2}}\neq 0$) and we
obtain\footnote{\label{foot:branch}Notice that the cancellation of the term
  that diverges in the $\rho\rightarrow 0$ limit can only be achieved in the
  branch in which $L_{0}>0$. In particular, if either $L_{+}<0$ or $L_{-}<0$
  we are forced to work in the $L_{0}>0$ branch and that contribution cannot
  be made to vanish,}

\begin{equation}
\begin{array}{rcl}
ds^{2} 
& = & 
\hat{f}^{2}dt^{2}-\hat{f}^{-1}(d\rho^{2}+\rho^{2}d\Omega_{(3)}^{2})\, ,
\\
& & \\
\hat{f}^{-3}
& = &
{\displaystyle
1
+
\frac{2e^{-\phi_{\infty}}k_{\infty}^{2/3}}{3g^{2}}
\frac{\rho^{2} +2\kappa^{2}}{(\rho^{2} +\kappa^{2})^{2}}
\, ,  
}
\\
& & \\
A^{0} 
& = &
{\displaystyle
-\frac{1}{\sqrt{3}B_{0}}\hat{f}^{3}dt\, ,
}
\hspace{1cm}
A^{A} 
=
{\displaystyle
\frac{\kappa^{2}}{g(\rho^{2} +\kappa^{2})}v_{R}^{A}\, ,  
}
\\
& & \\
e^{2\phi}
& = &
e^{2\phi_{\infty}}\hat{f}^{-3}\, ,
\hspace{1cm}
k
=
k_{\infty}\hat{f}^{3/4}\, ,
\end{array}
\end{equation}

This solution depends on one function, $\hat{f}$ which has the same profile as
the one appearing in the gauge 5-brane \cite{Strominger:1990et}.\footnote{More
  precisely, the function $H=e^{2\phi_{\infty}}\hat{f}^{-3}$.}  The
similarity can be made more manifest by using the relation between the
5-dimensional Yang-Mills coupling constant $g$, the Regge slope
$\alpha^{\prime}$, the string coupling constant $g_{s}=e^{\phi_{\infty}}$ and
the radius of compactification from 6 to 5 dimensions
$k_{\infty}=R_{z}/\ell_{s}$, where $\ell_{s}=\sqrt{\alpha^{\prime}}$ is the
string length parameter:

\begin{equation}
g = k^{1/3}_{\infty}e^{-\phi_{\infty}/2}/\sqrt{12\alpha^{\prime}}\, ,
\end{equation}

\noindent
which brings $e^{2\phi}$ to the form\footnote{In our conventions, which
  coincide essentially with those of Ref.~\cite{Strominger:1990et}, the
  10-dimensional Heterotic String effective action is written in the string
  frame as
 \begin{equation}
S_{\rm Het}
=
\frac{g_{s}^{2}}{16\pi G_{N}^{(10)}}
\int dx^{10} \sqrt{|g|}\, e^{-2\phi}
\left [R -4(\partial\phi)^{2} +\tfrac{1}{12}H^{2} 
-\alpha^{\prime}F^{A}F^{A}\right]\, .    
\end{equation}
The 10-dimensional string-frame metric solution is normalized such that it
becomes $(+1,-1,\cdots,-1)$ at spatial infinity. The same is true for the
5-dimensional metric, which can be seen as the modified-Einstein-frame metric
in the language of Ref.~\cite{Maldacena:1996ky}. The relation between these
two metrics involves rescalings by powers of $e^{\phi-\phi_{\infty}}$ and
$k/k_{\infty}$.}

\begin{equation}
e^{2\phi}
=
e^{2\phi_{\infty}}\hat{f}^{-3}
=
e^{2\phi_{\infty}}
\left\{1+
8\alpha^{\prime}
\frac{\rho^{2} +2\kappa^{2}}{(\rho^{2} +\kappa^{2})^{2}} \right\}\, .  
\end{equation}

It is not difficult to show that, indeed, this solution is nothing but the
double dimensional reduction of the gauge 5-brane compactified on $T^{5}$
\cite{Cano:2017qrq}.

From the purely 5-dimensional point of view, apart from the instanton field,
the solution has a vector field $A^{0}$ which is dual to the Kalb-Ramond
2-form and is sourced by the instanton number density only, as in the gauge
5-brane \cite{Duff:1991pe}. Observe that this means that the parameter $q_{0}$
is the sum of the instanton-number contributions (associated to a gauge
5-brane, as we are going to argue) which amount to just $\frac{2}{9g^{2}}$ and
electric sources of a different origin which amount to
$\tilde{q}_{0}=q_{0}-\frac{2}{9g^{2}}$ which we have set to zero in the above
solution. The complete identification of the higher-dimensional stringy
components of the general solution will be the subject of the forthcoming
paper \cite{Cano:2017qrq}. Here we just want to study the above solution, which in its
5-dimensional form is, apart from supersymmetric, clearly globally regular (at
least for finite values of $\kappa$), asymptotically flat and horizonless and
they are the higher-dimensional analogue of the \textit{global monopole}
solutions found in gauged $\mathcal{N}=4,d=4$ supergravity
\cite{Harvey:1991jr,Chamseddine:1997nm,Chamseddine:1997mc} and also in
$\mathcal{N}=2,d=4$ SEYM theories \cite{Hubscher:2008yz,Bueno:2014mea}.

The mass of the global instanton is obtained by replacing $q_{0}$ by
$\frac{2}{9g^{2}}$ and setting $q_{\pm}=0$ in  Eq.~(\ref{eq:mass}):

\begin{equation}
M
= 
\frac{\pi}{6 g^{2}G_{N}^{(5)}}  
e^{-\phi_{\infty}}k_{\infty}^{2/3}
= 
8\frac{R_{9}\cdots R_{5}}{g_{s}^{2}\ell_{s}^{6}}\, ,
\end{equation}

\noindent
where $R_{i}$ is the compactification radius of the $x^{i}$ coordinate and
where we have used 

\begin{equation}
G_{N}^{(5)}= \frac{G_{N}^{(10)}}{(2\pi)^{5}R_{9}\cdots R_{5}}\, ,
\,\,\,\,\,
\mbox{and}\,\,\,\,\,
G_{N}^{(10)} = 8\pi^{6}g_{s}^{2}\ell_{s}^{8}\, .
\end{equation}

\noindent
This value is eight times that of a single neutral (solitonic) 5-brane
\cite{Callan:1991ky,Callan:1991dj}.

The metric depends on the instanton scale $\kappa^{2}$, and it becomes
singular when $\kappa=0$. It is tempting to regard that singular metric as the
result of concentrating all the mass, which is independent of $\kappa$, in a
single point.  Thus, one may wonder how the radial distribution of the energy
depends on $\kappa$ and whether there is a value of $\kappa$ and $\rho$ such
that the energy enclosed in a 3-sphere of that radius is larger than the mass
of a Schwarzschild black hole of that Schwarzschild radius ($R_{S}^{2}=3\pi
M/(8G_{N}^{(5)})$).

\begin{figure}[t]
  \centering
  \includegraphics[height=7cm]{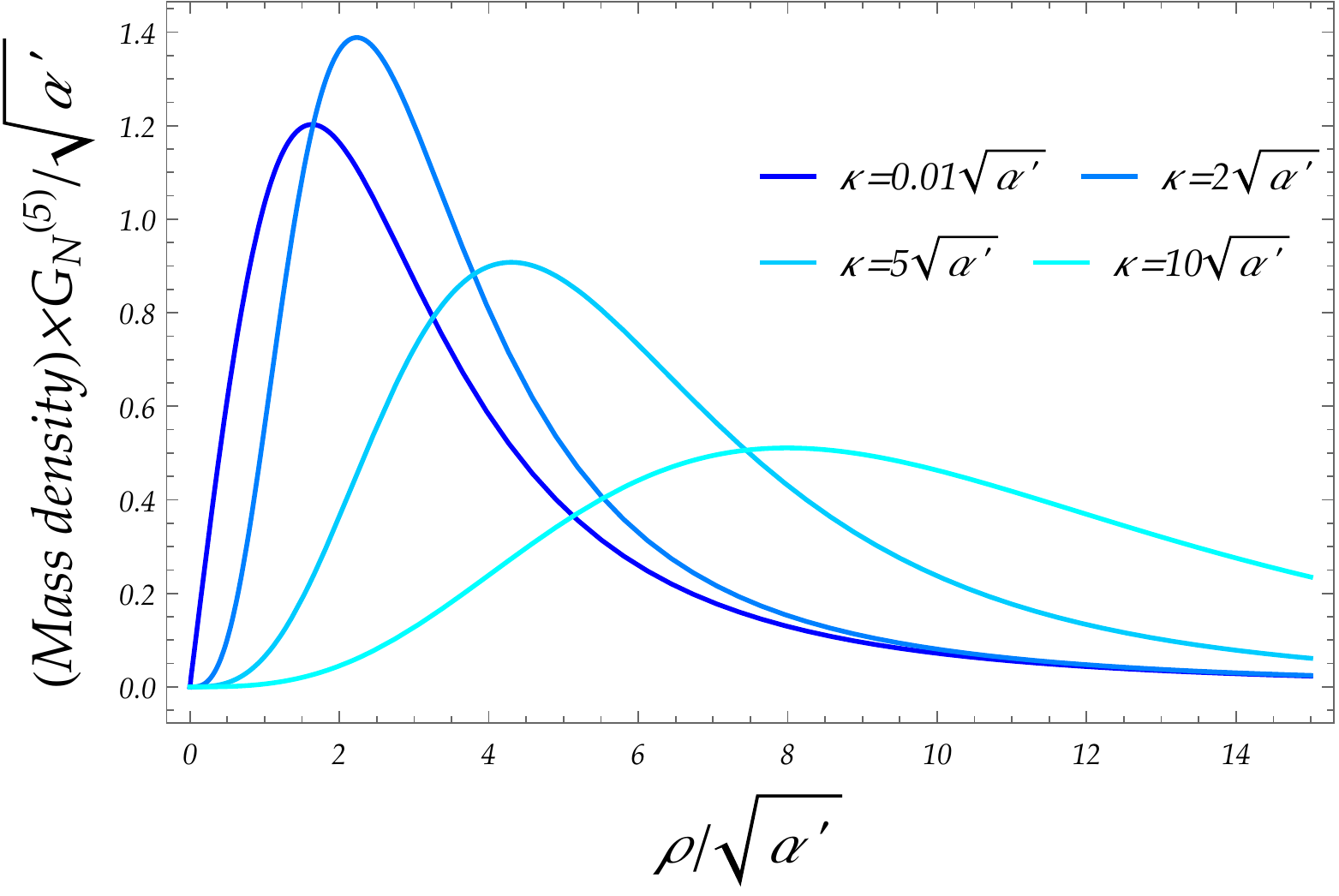}
  \caption{\small Radial mass density function of the global instanton solution for
    different values of the instanton scale, $\kappa^{2}$.}
  \label{fig:Massdensity}
\end{figure}

\begin{figure}[t]
  \centering
  \includegraphics[height=7cm]{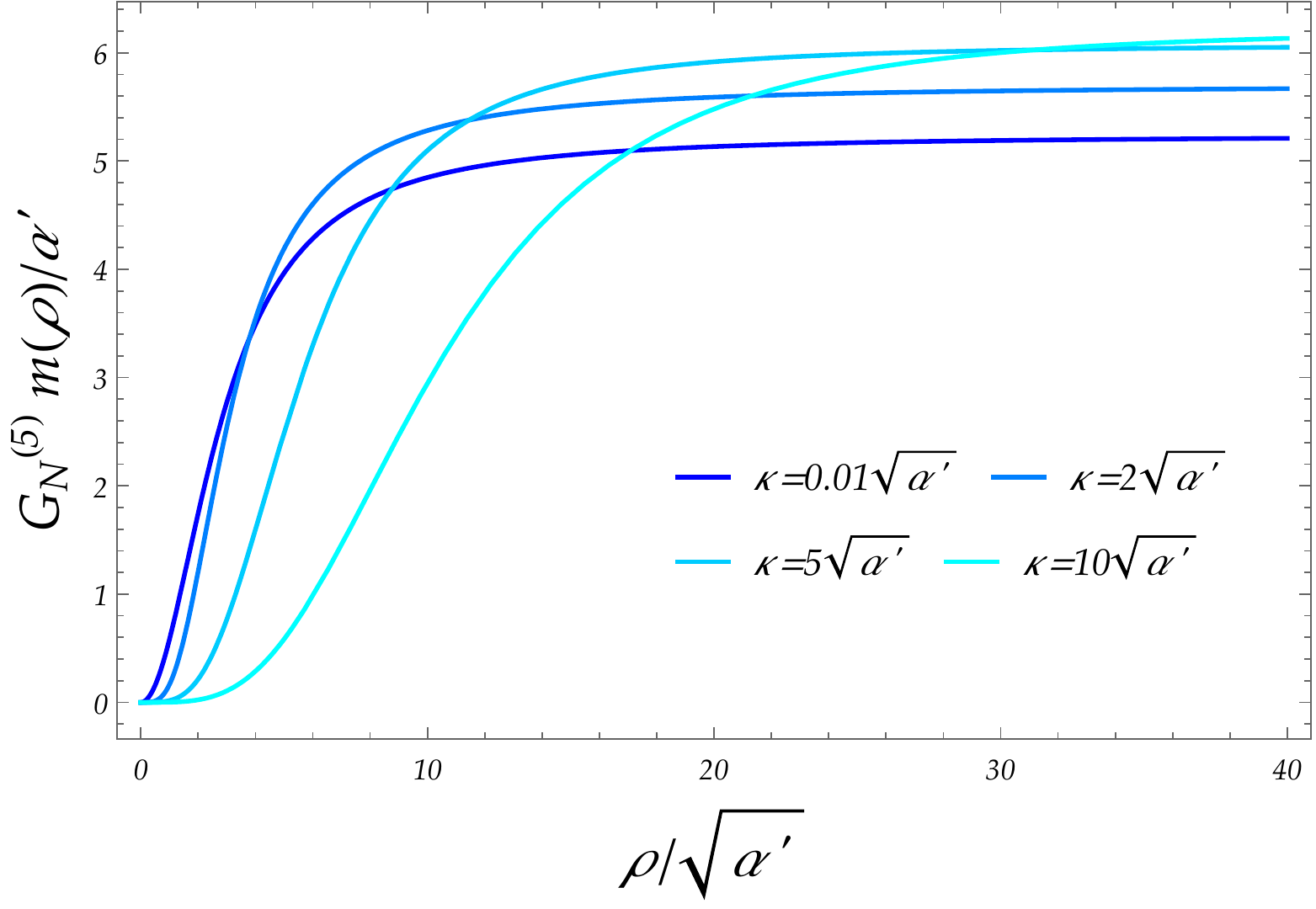}
  \caption{\small Radial mass function of the global instanton solution for different
    values of $\kappa^{2}$ obtained by integration of the mass
    density function in Fig.~\ref{fig:Massdensity} with respect to $\rho$.}
  \label{fig:Massfunction}
\end{figure}

\begin{figure}[t]
  \centering
  \includegraphics[height=7cm]{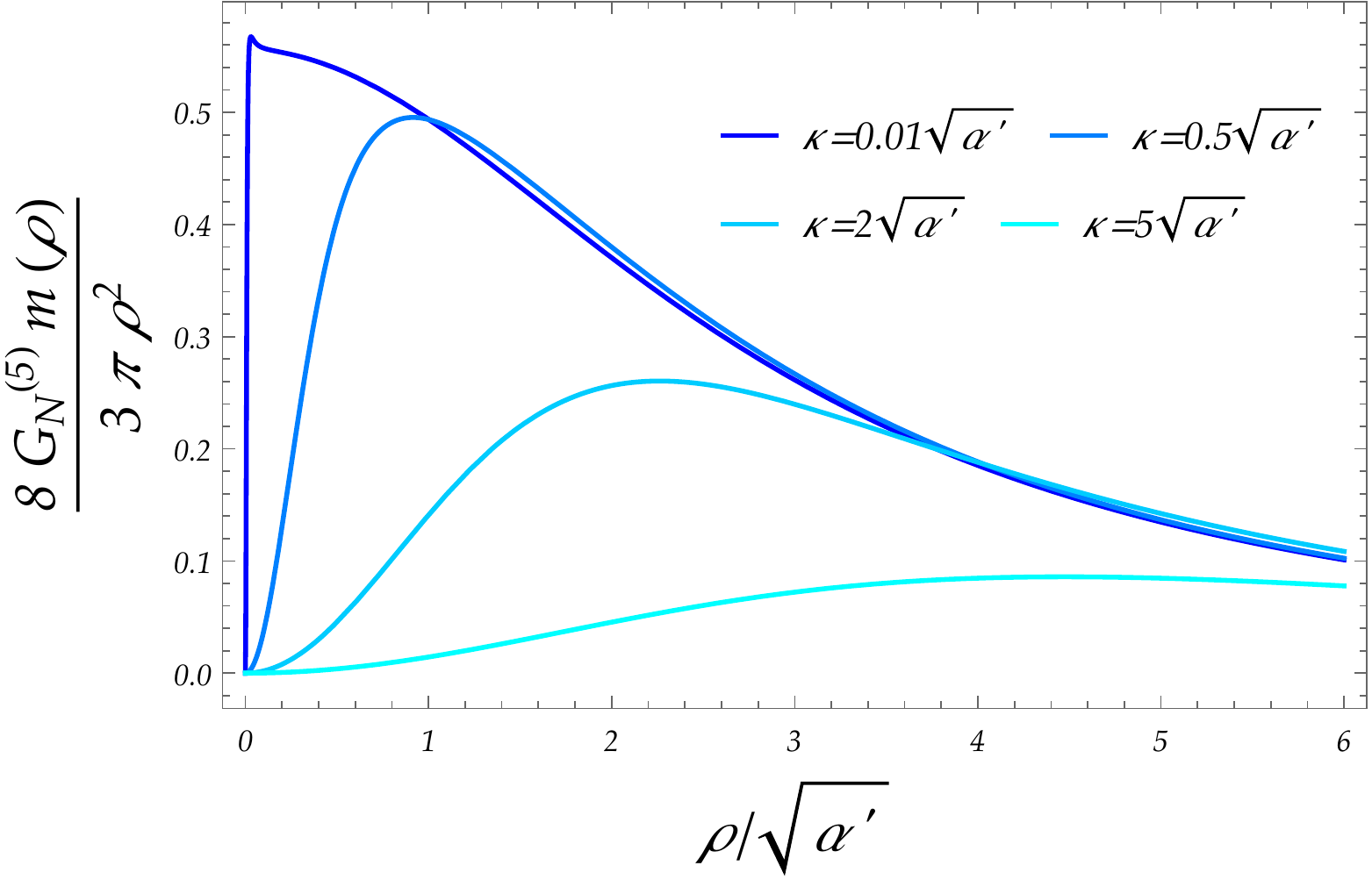}
  \caption{\small Quotient between the radial mass function of the global instanton
    solution and the mass of the 5-dimensional Schwarzschild black hole for
    that Schwarzschild radius for different values of $\kappa^{2}$.
  }
  \label{fig:Relativemass}
\end{figure}

The radial mass density, given by $\sqrt{|g|}T^{00}$
($T^{00}$ being the tangent-space basis component of the energy-momentum
tensor) is represented in Fig.~\ref{fig:Massdensity} for different values of
the instanton scale and its integral over a sphere of radius $R$ (the
\textit{mass function}) is represented in Fig.~\ref{fig:Massfunction}. The
values of the integrals at infinity are not exactly equal because, after all,
there is no well-defined concept of energy density in General Relativity and
we are just using a reasonable approximation. In Fig.~\ref{fig:Relativemass}
we have represented the quotient between the mass function and the
Schwarzschild mass as a function of $R$ and we see that it never goes above
$5/9$ for any finite, non-vanishing value of the instanton scale.

\begin{figure}[t]
  \centering
  \includegraphics[height=7cm]{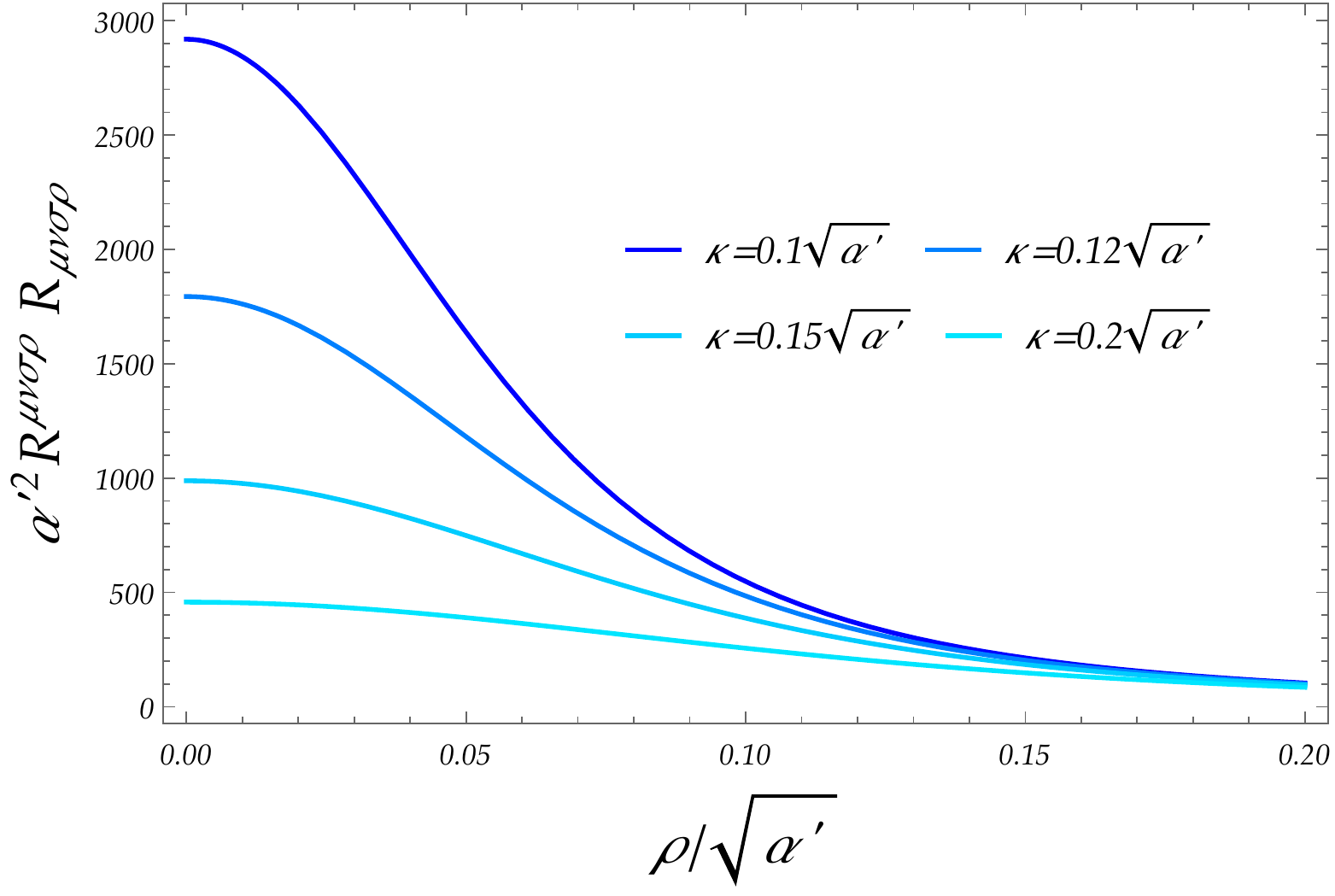}
  \caption{\small Value of the Kretschmann invariant for the global instanton
    solution for different values of $\kappa^{2}$.}
  \label{fig:Kretschmann}
\end{figure}

\section{Conclusions}
\label{sec-conclusions}

Globally regular solutions supported by elementary fields are quite
remarkable. In the case of the 4-dimensional global monopoles
\cite{Harvey:1991jr,Chamseddine:1997nm,Chamseddine:1997mc,Hubscher:2008yz,Bueno:2014mea}
we have argued that they represent elementary, non-perturbative states of the
theory because they do not modify the entropy of a given Abelian black hole
solution when they are added to it. They do contribute to the mass,
though. Adding the global instanton to 5-dimensional black holes should have
the same result: unmodified entropy and increased mass. However, the reverse
seems to happen: the entropy is modified while the mass is not. The
construction of the global instanton solution seems to suggest that this is a
false appearance caused by an inappropriate definition of the charges
involved. The exact role in 4-dimensional non-Abelian black-hole solutions (in
which it must appear disguised as a \textit{coloured monopole}) has to be
investigated. It is also unclear if a global instanton can be added to a
Schwarzschild-Tangerlini (or any other non-extremal black hole) and what the
effect would be.

We have tried to deform this solution by adding angular momentum, which in
these theories is always possible, although the simplest ways to do it (adding
a non-trivial harmonic function $M$ to generate a non-vanishing $\omega_{5}$)
would also introduce a singularity at the origin. While we have succeeded in
producing an $\omega_{5}$ regular at $\rho=0$ and dropping at infinity as
$\rho^{-2}$, the metric function $\hat{f}^{-1}$ becomes singular at
$\rho=0$. It is possible to cancel those singularities by introducing
additional Abelian harmonic functions with fine-tuned coefficients but the
resulting $\hat{f}^{-1}$ either has zeroes, or leads to negative mass or both.

The non-Abelian solutions found so far in the supergravity/superstring context
are the simplest to construct. One can expect, however, a space of solutions
far richer than that of the Abelian ones. Work in this direction is under way.

\section*{Acknowledgments}

The authors would like to thank P.~Meessen and C.S.~Shahbazi for interesting
conversations.  This work has been supported in part by the Spanish Government
grants FPA2012-35043-C02-01 and FPA2015-66793-P (MINECO/FEDER, UE), the Centro
de Excelencia Severo Ochoa Program grant SEV-2012-0249 and the Spanish
Consolider-Ingenio 2010 program CPAN CSD2007-00042.  The work of PAC was
supported by Fundaci\'on la Caixa under ``la Caixa-Severo Ochoa''
International pre-doctoral grant. The work of PFR was supported by the
\textit{Severo Ochoa} pre-doctoral grant SVP-2013-067903.  TO wishes to thank
M.M.~Fern\'andez for her permanent support.

\appendix

\section{The theory}
\label{sec-thetheory}

The theory we are considering is a truncation of the effective field theory of
the Heterotic Superstring compactified on $T^{5}$ that preserves an SU$(2)$
triplet of vector fields. The compactification and truncation reduce the
theory to a particular model of gauged $\mathcal{N}=1,d=5$ supergravity to
which one can apply the solution-generating techniques based on the
characterization of supersymmetric solutions described in
Appendix~\ref{sec-susysolutions}. The dimensional reduction of this model on a
circle gives the so-called ST$[2,6]$ model of $\mathcal{N}=2,d=4$ supergravity
coupled to 6 vector multiplets and we will, therefore, refer to it by that
name in the 5-dimensional context as well.  Here we are going to give a
minimal description of the bosonic sector of these theories and of the
particular model we are considering. More information can be found in
Refs.~\cite{Bergshoeff:2004kh,Freedman:2012zz,Ortin:2015hya}.\footnote{Our
  conventions are those in
  Refs.~\cite{Bellorin:2006yr,Bellorin:2007yp,Ortin:2015hya} which are those
  of Ref.~\cite{Bergshoeff:2004kh} with minor modifications.}

The ST$[2,6]$ model model $\mathcal{N}=1,d=5$ supergravity contains 5 vector
supermultiplets labeled by $x,y=1,\cdots,5$, each containing a vector field
$A^{x}{}_{\mu}$ and a scalar $\phi^{x}$. Together with the graviphoton
$A^{0}{}_{\mu}$, all the vectors are written $A^{I}{}_{\mu}$, $I,J,\ldots=
0,1,\cdots,5$. The only remaining bosonic field is the spacetime metric
$g_{\mu\nu}$.  The $C_{IJK}$ tensor has the non-vanishing components

\begin{equation}
C_{0xy}= \tfrac{1}{6}\eta_{xy}\, , \qquad
\mbox{where}
\,\,\,\,\,
(\eta_{xy}) = \mathrm{diag}(+-\dotsm -)\, ,
\end{equation}

\noindent
and the Real Special manifold parametrized by the physical scalars can be
identified with the Riemannian symmetric space

\begin{equation}
\mathrm{SO}(1,1)\times \frac{\mathrm{SO}(1,4)}{\mathrm{SO}(4)}\, .
\end{equation}

\noindent
A convenient parametrization of the scalar manifold is 

\begin{equation}
\label{eq:parametrization}
h^{0} 
= 
e^{-\phi}k^{2/3}\, ,
\hspace{.3cm}
h^{1,2} 
=
k^{-4/3}\left[1\pm (\ell^{2} +\tfrac{1}{2}e^{\phi}k^{2}) \right] \, ,
\hspace{.3cm}
h^{3,4,5} = -2k^{-4/3} \ell^{3,4,5}\, ,
\end{equation}

\noindent
where $\phi$ coincides with the 10-dimensional Heterotic Superstring dilaton
field, $k$ is the Kaluza-Klein scalar of the dimensional reduction from $d=6$
to $d=5$ and the $\ell^{A}$ are the fifth components of the 6-dimensional
vector fields. The rest of the components that make up the 10-dimensional
vector fields have been truncated \cite{Cano:2016rls}.



The group SO$(3)$ acts in the adjoint on the coordinates $x=3,4,5$ which we
are going to denote by $A,B,\ldots$ and this is the sector that is gauged
without the use of Fayet-Iliopoulos terms. This means that R-symmetry is not
gauged and there is no scalar potential.\footnote{Models of this kind are
  called model of $\mathcal{N}=1,d=5$ Super-Einstein-Yang-Mills (SEYM), which
  are the simplest $\mathcal{N}=1$ supersymmetrization of the 5-dimensional
  Einstein-Yang-Mills (EYM) theories.} The structure constants are
$f_{AB}{}^{C} =+\varepsilon_{AB}{}^{C}$.\footnote{These indices will always be
  raised and lowered with $\delta_{AB}$, just for esthetical reasons.}  We
will denote with $a,b,\ldots = 1,2$ the ungauged directions.  Observe that
this sector of the theory corresponds to the so-called STU model: in absence
of the $h^{A}$s we can make the linear redefinitions

\begin{equation}
h^{1\prime} \equiv \tfrac{1}{\sqrt{2}}(h^{1}+h^{2})\, ,
\hspace{1cm}
h^{2\prime} \equiv \tfrac{1}{\sqrt{2}}(h^{1}-h^{2})\, ,
\,\,\,\,
\Rightarrow
\,\,\,\,
C_{abc}h^{a}h^{b}h^{c}
=
h^{0}h^{1\prime} h^{2\prime}\, .  
\end{equation}

\noindent
Thus, our model can be also understood as the STU model with an additional
SU$(2)$ triplet of vector multiplets. 

With the above parametrization of the scalar manifold, the action for this
model can be brought to the form

\begin{equation}
\label{eq:gaugedST[2,6]action}
\begin{array}{rcl}
S 
& = &  
{\displaystyle\int 
d^{5}x\sqrt{g}\
\biggl\{
R
+\partial_{\mu}\phi\partial^{\mu}\phi
+\tfrac{4}{3}\partial_{\mu}\log{k}\partial^{\mu}\log{k}
+2e^{-\phi}k^{-2}\mathfrak{D}_{\mu}\ell^{A}
\mathfrak{D}^{\mu}\ell^{A}
}
\\ \\ & & 
-\tfrac{1}{12} e^{2\phi} k^{-4/3} F^{0}\cdot F^{0}
+\tfrac{1}{12}\left(\eta_{xy}e^{-\phi}k^{2/3}
-9h_{x}h_{y}\right)
F^{x}\cdot F^{y}
\\ \\ & & 
+\tfrac{1}{24\sqrt{3}}
{\displaystyle\frac{\varepsilon^{\mu\nu\rho\sigma\alpha}}{\sqrt{g}}}
A^{0}{}_{\mu} \eta_{xy}F^{x}{}_{\nu\rho}F^{y}{}_{\sigma\alpha}
\biggr\}\, ,
\end{array}
\end{equation}

\noindent
where

\begin{eqnarray}
\mathfrak{D}_{\mu}\ell^{A}
& = & 
\partial_{\mu} \ell^{A}
+g\varepsilon^{A}{}_{BC} A^{B}{}_{\mu}\ell^{C}\, ,
\\
& & \nonumber \\
F^{0,a}{}_{\mu\nu}
& = &
2\partial_{[\mu}A^{0,a}{}_{\nu]}\, ,
\\
& & \nonumber \\
F^{A}{}_{\mu\nu}
& = &
2\partial_{[\mu}A^{A}{}_{\nu]}  
+g\varepsilon^{A}{}_{BC}A^{B}{}_{\mu}A^{C}{}_{\nu}\, .
\end{eqnarray}

Notice that $A^{0}{}_{\mu}$ is sourced by
$\varepsilon^{\mu\nu\rho\sigma\alpha}\eta_{xy}F^{x}{}_{\nu\rho}F^{y}{}_{\sigma\alpha}$
which is related to the instanton number on the constant-time
hypersurfaces. In differential-form language, its equation of motion is

\begin{equation}
d(e^{2\phi}k^{-4/3}\star F^{0})= \tfrac{1}{2\sqrt{3}} \eta_{xy}F^{x}\wedge F^{y} =0\, ,
\end{equation}

\noindent
which is similar to that of the Kalb-Ramond 2-form $B$. This is because
$A^{0}$ is the 5-dimensional dual of the dimensionally reduced Heterotic
Kalb-Ramond form $B$. The duality relation is

\begin{equation}
F^{0}= e^{-2\phi}k^{4/3} \star H\, ,  
\,\,\,\,\,
\mbox{with}
\,\,\,\,\,
H \equiv  dB +\tfrac{1}{2\sqrt{3}}\omega_{\rm CS}\, ,
\end{equation}

\noindent
where $\omega_{\rm CS}$ is the Chern-Simons 3-form of all the vector fields
but $A^{0}$ itself

\begin{equation}
\omega_{\rm CS}
= 
\tfrac{1}{2}F^{+}\wedge A^{-} + \tfrac{1}{2}F^{-}\wedge A^{}
+
F^{A}\wedge A^{A} -\tfrac{1}{3!}g \epsilon_{ABC}A^{A}\wedge A^{B}\wedge
A^{C}\, ,   
\end{equation}

\noindent
satisfying

\begin{equation}
d\omega_{\rm CS} = \eta_{xy}F^{x}\wedge F^{y}\, . 
\end{equation}

\section{Timelike supersymmetric solutions}
\label{sec-susysolutions}

As shown in
Refs.~\cite{Huebscher:2007hj,Hubscher:2008yz,Bellorin:2007yp,Bellorin:2008we,Meessen:2015enl},
the problem of finding timelike supersymmetric solutions of
$\mathcal{N}=2,d=4$ SEYM theories and timelike or null supersymmetric
solutions with an additional isometry of $\mathcal{N}=1,d=5$ SEYM theories is
effectively reduced to a much simpler problem: finding functions
$\Phi^{\Lambda},\Phi_{\Lambda}$ and vector fields
$\breve{A}^{\Lambda}{}_{\underline{r}}$\footnote{$\Lambda,\Sigma,\ldots=0,1,\cdots,n_{V5}+1$
  where $n_{V5}$ is the number of vector supermultiplets in $d=5$ and
  $r,s,\ldots=1,2,3$.} in Euclidean 3-dimensional space $\mathbb{E}^{3}$
solving these three sets of equations:

\begin{eqnarray}
\label{eq:B}
\tfrac{1}{2}\varepsilon_{\underline{r}\underline{s}\underline{w}}
\breve{F}^{\Lambda}{}_{\underline{s}\underline{w}}
-\breve{\mathfrak{D}}_{\underline{r}}\Phi^{\Lambda}
& = &
0\, ,
\\
& & \nonumber \\
\label{eq:D2Phi}
\breve{\mathfrak{D}}_{\underline{r}}\breve{\mathfrak{D}}_{\underline{r}}\Phi_{\Lambda} 
-g^{2} f_{\Lambda\Sigma}{}^{\Omega}f_{\Delta\Omega}{}^{\Gamma}
\Phi^{\Sigma}\Phi^{\Delta}\Phi_{\Gamma}
& = &
0\, ,
\\
& & \nonumber \\
\label{eq:integrability}
\Phi_{\Lambda}\breve{\mathfrak{D}}_{\underline{r}}\breve{\mathfrak{D}}_{\underline{r}}\Phi^{\Lambda}
-
\Phi^{\Lambda}\breve{\mathfrak{D}}_{\underline{r}}\breve{\mathfrak{D}}_{\underline{r}}\Phi_{\Lambda}
& = &
0\, ,
\end{eqnarray}

\noindent
where $\breve{\mathfrak{D}}_{\underline{r}}$ is the gauge covariant derivative
in $\mathbb{E}^{3}$ with respect to the connection
$\breve{A}^{\Lambda}{}_{\underline{r}}$.

Eqs.~(\ref{eq:B}) are the Bogomol'nyi equations \cite{Bogomolny:1975de} for a
set of real, adjoint, Higgs fields $\Phi^{\Lambda}$ and gauge vector fields
$\breve{A}^{\Lambda}{}_{\underline{r}}$ on $\mathbb{E}^{3}$. In the Abelian
case, the integrability conditions are the Laplace equations
$\partial_{\underline{r}}\partial_{\underline{r}}\Phi^{\Lambda}=0$ and the
vector fields are implicitly determined by the harmonic functions
$\Phi^{\Lambda}$. In the non-Abelian sector this is no longer true, and the non-linear equation has to be solved simultaneously for the scalar and the vector fields. 

Eqs.~(\ref{eq:D2Phi}) are equations for the scalar fields $\Phi_{\Lambda}$
linear in them. In the Abelian directions the $\Phi_{\Lambda}$ are harmonic
functions
$\partial_{\underline{r}}\partial_{\underline{r}}\Phi_{\Lambda}=0$. In the
SU$(2)$ directions we are going to set them to zero.\footnote{Non-trivial
  solutions are also available: for any compact group one can take
  $\Phi_{\Lambda} = K \Phi^{\Lambda}$ for some constant $K$ and, for SU$(2)$
  more interesting solutions have been recently found in
  Ref.~\cite{Ramirez:2016tqc} using the results of
  Refs.~\cite{Etesi:2002cc,kn:KronheimerMScThesis}, but they are only relevant
  in multicenter solutions \cite{MOR}.}

Eq.~(\ref{eq:integrability}) is the integrability condition of the equations
that define the 1-forms $\omega_{\underline{r}}$ that appear in the 4- and
5-dimensional metrics 

\begin{equation}
\label{eq:omega}
\partial_{[\underline{r}}\omega_{\underline{s}]} 
=
2\varepsilon_{rsw}
\left(
\Phi_{\Lambda} \breve{\mathfrak{D}}_{\underline{w}}\Phi^{\Lambda}
-
\Phi^{\Lambda} \breve{\mathfrak{D}}_{\underline{w}}\Phi_{\Lambda}
\right)\, .
\end{equation}

\noindent
and it is guaranteed to be satisfied everywhere except
at the loci of the singularities of the scalar functions
$\Phi^{\Lambda},\Phi_{\Lambda}$ where it lead to the so-called \textit{bubble
  equations}\footnote{ See
  Refs.~\cite{Denef:2000nb,Bates:2003vx,Bena:2007kg}}.

For each solution
$\Phi^{\Lambda},\Phi_{\Lambda},\breve{A}^{\Lambda}{}_{\underline{r}}$ we can
construct two different solutions of the three kinds mentioned above. Here we
only need the prescription to construct timelike solutions with an additional
isometry of $\mathcal{N}=1,d=5$ SEYM theories:

\begin{enumerate}
\item The elementary building blocks, namely the $2(n_{V5}+2)$ functions
  $M,H,K^{I},L_{I}$ and the 1-forms $\omega,\breve{A}^{I},\chi$ in
  $\mathbb{E}^{3}$ are related to the functions
  $\Phi^{\Lambda},\Phi_{\Lambda}$ and 1-forms
  $\omega,\breve{A}^{\Lambda}{}_{\underline{r}}$ determined by solving
  Eqs.~(\ref{eq:B})-(\ref{eq:omega}) by 

\begin{equation}
\label{eq:5dtimelikeidentifications}
\begin{array}{rcl}
K^{I} & = & \delta^{I}{}_{\Lambda}\Phi^{\Lambda+1}\, ,
\hspace{.7cm}
L_{I} = -\frac{2\sqrt{2}}{3}\delta_{I}{}^{\Lambda}\Phi_{\Lambda+1}\, ,  
\hspace{.7cm}
H = -2\sqrt{2}\Phi^{0}\, ,
\hspace{.7cm}
M = +\sqrt{2}\Phi_{0}\, ,
\\
& & \\
\omega & = & \omega\, ,
\hspace{.7cm}
\chi_{\underline{r}} = -2\sqrt{2}\breve{A}^{0}{}_{\underline{r}}\, ,  
\hspace{.7cm}
\breve{A}^{I}{}_{\underline{r}}
= 
\delta^{I}{}_{\Lambda}\breve{A}^{\Lambda+1}{}_{\underline{r}}\, ,
\,\,\,\,\,
I=0,\cdots,n_{V5}\, .
\end{array}
\end{equation}

All the timelike solutions have necessarily $H\neq 0$, ($\Phi^{0}\neq 0$).

\item The Yang-Mills coupling constant that appears in
  Eqs.~(\ref{eq:B})-(\ref{eq:integrability}) can be understood as the
  4($=1+3$)-dimensional one $g_{4}$. It needs to be replaced by the coupling
  constant used in the 5-dimensional theory, which is related to it by

  \begin{equation}
  g_{4}= -\sqrt{24}g\, .  
  \end{equation}

\item With the above building blocks we construct first the combinations

\begin{eqnarray}
\label{eq:hIf}
h_{I}/\hat{f} 
& = &
L_{I}+8C_{IJK}K^{J}K^{K}/H\, ,
\\
& & \nonumber \\
\hat{\omega} 
& = & 
\omega_{5}(dz+\chi) +\omega\, ,
\\
& & \nonumber \\
\label{eq:omega5}
\omega_{5}
& = &
M+16\sqrt{2} H^{-2} C_{IJK} K^{I} K^{J} K^{K}
+3\sqrt{2} H^{-1} L_{I} K^{I} \, ,
\\
& & \nonumber \\
\label{eq:Kromheimer}
\hat{A}^{I}
& = &
2\sqrt{6} \left[H^{-1}K^{I} (dz+\chi)-\breve{A}^{I} \right]\, ,
\\
& & \nonumber \\
\hat{F}^{I}
& = &
2\sqrt{6} 
\left\{\breve{\mathfrak{D}} \left[K^{I} H^{-1} \wedge (dz+\chi)\right]
-\star_{3} H \breve{\mathfrak{D}} K^{I} \right\} \, .
\end{eqnarray}

\item The physical fields are recovered from the building blocks as follows:

\begin{enumerate}
\item For Real Special manifolds which are Riemannian symmetric manifolds we
  can use this expression

\begin{equation}
\label{eq:f3symmetric-2}
\begin{array}{rcl}
\hat{f}^{-3}
& = &
3^{3} C^{IJK}L_{I}L_{J}L_{K}
+3^{4}\cdot 2^{3}  C^{IJK}C_{KLM}L_{I}L_{J}K^{L}K^{M}/H
\\
& & \\
& &
+3\cdot 2^{6}L_{I}K^{I}C_{JKL}K^{J}K^{K}K^{L}/H^{2}
+2^{9}\left(C_{IJK}K^{I}K^{J}K^{K}\right)^{2}/H^{3}\, ,
\end{array}
\end{equation}

\noindent
which for the model at hands reduces to

\begin{equation}
\label{eq:d5metricfunctionST24}
  \begin{array}{rcl}
\hat{f}^{\, -1}
& = &
H^{-1}
\left\{
\tfrac{1}{4}
\left(
6HL_{0} +8\eta_{xy}K^{x}K^{y}
\right)
\left[
9H^{2}\eta^{xy}L_{x}L_{y} +48 HK^{0}L_{x}K^{x}
\right.
\right.
\\
& & \\
& & 
\left.\left.
+64(K^{0})^{2}\eta_{xy}K^{x}K^{y}
\right]
\right\}^{1/3}\, .
\end{array}
\end{equation}

\item Using the metric factor we can find the $h_{I}$ from Eq.~(\ref{eq:hIf})
  and, from these, the $h^{I}$ using 

\begin{equation}
h^{I} = 27 C^{IJK}h_{J}h_{K}\, .
\end{equation}

The scalar fields $\phi^{x}$ can be obtained by inverting the functions
$h_{I}(\phi)$ or $h^{I}(\phi)$. A possible, but not unique, parametrization can be given by

\begin{equation}
\phi^{x}=h^{x}= 9\eta^{xy}h_{y}h_{0} \, .
\end{equation}

\item With the previous results the spacetime metric is completely determined
  and has the form

\begin{equation}
ds^{2} = \hat{f}^{\, 2}(dt+\hat{\omega})^{2}
-\hat{f}^{\, -1}\left[H^{-1}(dz+\chi)^{2} +H dx^{r} dx^{r}\right]\, .  
\end{equation}

\item The 5-dimensional vector fields are given by

\begin{equation}
\label{eq:completevectorfields}
A^{I} = -\sqrt{3}h^{I}e^{0} +\hat{A}^{I}\, ,    
\,\,\,\,\,
\mbox{where}
\,\,\,\,\,
e^{0} 
\equiv
\hat{f} (dt +\hat{\omega})\, ,
\end{equation}

\noindent
so that the spatial components, labeled by $m,n=z,1,2,3$, are

\begin{equation}
A^{I}{}_{\underline{m}} 
= 
\hat{A}^{I}{}_{\underline{m}} -\sqrt{3}h^{I}\hat{f} \hat{\omega}_{\underline{m}}\, .  
\end{equation}




\end{enumerate}

\end{enumerate}


\end{document}